\documentclass[11pt]{article}
\usepackage{graphics}
\usepackage{epsfig}
\catcode`\@=11

\def\beqra{\begin{eqnarray}} 
\def\eeqra{\end{eqnarray}}

\newcommand{\be}[1]{\begin{equation}\label{#1}}
\newcommand{\beq}{\begin{equation}}
\newcommand{\ee}{\end{equation}}
\newcommand{\beqn}[1]{\begin{eqnarray}\label{#1}}
\newcommand{\eeqn}{\end{eqnarray}}

\def\rhb{\bar{\rho}}

\def\bg{\bar{g}_\ast}

\def\eps{\epsilon}

\def\Ga{\Gamma}

\def\DN{\Delta N_\nu}

\def\ifmath#1{\relax\ifmmode #1\else $#1$\fi}

\def\bold#1{\setbox0=\hbox{$#1$}%
     \kern-.025em\copy0\kern-\wd0
     \kern.05em\copy0\kern-\wd0
     \kern-.025em\raise.0433em\box0 }

\def\GENITEM#1;#2{\par\vskip6pt \hangafter=0 \hangindent=#1
   \Textindent{$ #2$ }\ignorespaces}

\def\mst11{m_{\;\widetilde{t}_{1}}}

\def\mst22{m_{\;\widetilde{t}_{2}}}
\def\mst12{m_{\;\widetilde{t}_{1,2}}}

\def\msb11{m_{\;\widetilde{b}_{1}}}
\def\msb22{m_{\;\widetilde{b}_{2}}}
\def\msb12{m_{\;\widetilde{b}_{1,2}}}

\def\mwidetilde2{\widetilde{m}^{2}}

\def\lesssim{\mathrel{\mathop
  {\hbox{\lower0.5ex\hbox{$\sim$}\kern-0.8em\lower-0.7ex\hbox{$<$}}}}}
\def\greatsim{\mathrel{\mathop
  {\hbox{\lower0.5ex\hbox{$\sim$}\kern-0.8em\lower-0.7ex\hbox{$>$}}}}}
\relax

\begin{document}

\begin{titlepage}

\vspace*{-64pt}
\begin{flushright}
DFAQ-TH/2000-03\\
INFNFE-04-00 \\ 
hep-ph/0008105 \\ 
July 2000 
\end{flushright}

\vskip .7cm

\begin{center}
{\Large \bf  The Early Mirror Universe: Inflation, \\
\vspace{0.3cm}
Baryogenesis, Nucleosynthesis  and  Dark Matter }
\vskip .7cm

{\bf Zurab Berezhiani,$^{a,b,}$\footnote{E-mail:
     {\tt berezhiani@fe.infn.it, berezhiani@aquila.infn.it}}}  
{\bf Denis Comelli$^{c,}$\footnote{E-mail:
     {\tt  comelli@fe.infn.it}}}  and 
{\bf Francesco L. Villante $^{c,d,}$\footnote{E-mail:
     {\tt villante@fe.infn.it}}}

\vskip.35in
{\it 
$^a$ Dipartimento di Fisica, Universit\`a di L'Aquila, 67010 Coppito  
AQ, and  \\ 
INFN, Laboratori Nazionali del Gran Sasso, 67010 Assergi AQ, 
Italy \\ 
$^b$ Andronikashvili Institute of Physics, Georgian Academy of
Sciences, \\ 
380077 Tbilisi, Georgia \\ 
$^c$ INFN, sezione di Ferrara, 44100 Ferrara, Italy\\
$^d$ Dipartimento di Fisica, Universit\`a di Ferrara, 41000 Ferrara, 
Italy \\
\vspace{12pt}
}

\end{center}

\vskip .5cm

\begin{abstract}

There can exist a parallel `mirror' world 
which has the same particle physics as the observable world 
and couples the latter only gravitationally.  
The nucleosynthesis bounds demand that the mirror sector 
should have a smaller temperature than the ordinary one.  
By this reason its evolution should be 
substantially deviated from the standard cosmology 
as far as the crucial epochs like 
baryogenesis, nucleosynthesis etc. are concerned. 
Starting from an inflationary scenario which could explain 
the different initial temperatures of the two sectors,
we study the time history of the early mirror universe. 
In particular, we show that in the context of the GUT or 
electroweak baryogenesis scenarios, 
the baryon asymmetry in the mirror world should be larger 
than in the observable one and in fact the mirror baryons 
could provide the dominant dark matter component 
of the universe. 
In addition, analyzing the nucleosynthesis epoch, 
we show that the mirror helium abundance should be much  
larger than that of ordinary helium.
The implications of the mirror baryons representing 
a kind of self-interacting dark matter 
for the large scale structure formation, the CMB anysotropy, 
the galactic halo structures, microlensing, etc. are briefly 
discussed. 

\end{abstract}

\end{titlepage}

\setcounter{footnote}{0}
\setcounter{page}{1}
\newpage
%

\section{Introduction}

The old idea 
that there can exist a hidden mirror sector of 
particles and interactions which is the exact duplicate 
of our visible world \cite{LY} 
has attracted a significant interest over last years, 
in particular being motivated by the problems of 
neutrino physics \cite{FV,BM}, gravitational microlensing  
\cite{BDM,Macho}, gamma ray bursts \cite{GRB}, 
ultra-high energy cosmic rays \cite{Venya}, 
flavour and CP violation \cite{PLB98}, etc. 
The basic concept is to have a theory given by the 
product $G\times G'$ of two identical gauge factors with 
the identical particle contents, which could naturally emerge 
e.g. in the context of $E_8\times E'_8$ superstring.
Two sectors communicate through gravity and  
perhaps also via some other messengers.  
A discrete symmetry $P(G\leftrightarrow G')$ interchanging
corresponding fields of $G$ and $G'$, so called mirror parity,  
implies that both particle sectors are described by the 
same Lagrangians.\footnote{In the brane world picture, 
the M-sector can be the same O-world realized on a 
parallel brane, $G'=G$ \cite{Manyfold}. 
}

In particular, one can consider 
a minimal symmetry $G_{SM}\times G'_{SM}$  
where $G_{SM}=SU(3)\times  SU(2)\times U(1)$ stands for the
standard model of observable particles: three families of 
quarks and leptons $q_i, ~u^c_i, ~d^c_i; ~l_i, ~e^c_i$
($i=1,2,3$) and the Higgs doublet $\phi$, 
while $G'_{SM}=[SU(3)\times SU(2)\times U(1)]'$ is its mirror
gauge counterpart with analogous particle content:
fermions $q'_i, ~u'^c_i, ~d'^c_i; ~l'_i, ~e'^c_i$  and 
the Higgs $\phi'$. 
(From now on all fields and quantities of the 
mirror (M) sector will have an apex 
to distinguish from the ones belonging to the 
observable or ordinary (O) world.) 
The mirror parity 
implies that all coupling constants (gauge, Yukawa, Higgs) 
have the same pattern in both sectors and thus  
their microphysics is the same.\footnote{ The mirror parity 
could be spontaneously broken and the weak interaction scales
$\langle \phi \rangle =v$ and $\langle \phi' \rangle =v'$
could be different, which leads to somewhat different particle 
physics in the mirror sector \cite{BM,BDM,Venya}. } 

One could naively think that due to mirror parity the 
O- and M- particles should have the same cosmological densities,  
which would be in the immediate conflict with 
the Big Bang nucleosynthesis (BBN) bounds 
on  the effective number of extra light neutrinos,  
$\DN <1$ \cite{io}:  
the mirror photons, electrons and neutrinos 
would give a contribution to the Hubble expansion 
rate equivalent to $\DN\simeq 6.14$.    
Therefore, the M-particle density in the early 
universe should be appropriately reduced. 
This situation is plausible if two following conditions 
are satisfied: 

A. At the initial moment  
two systems are born with different densities.   
In particular, the inflationary reheating temperature in 
the M-sector should be lower than in the visible one,   
$T'_R < T_R$, which can be achieved in certain models 
\cite{BDM,Venya,KST}. 

B. The M- and O-particles interact very weakly, 
so that two systems do not come into the thermal equilibrium 
with each other in the early universe.  
This condition is automatically fulfilled  if two worlds 
communicate only via the gravity. 
More generally,  there could be other messengers like  
superheavy gauge singlet fields 
or light singlets of the moduli type. 
In either case, they should mediate the 
effective couplings between the O- and M- particles 
suppressed by a large mass factor $M\sim M_{P}$ or so.

If two sectors have different reheating temperatures, and 
if they do not come into the thermal contact at later stages, 
then during the universe expansion they evolve independently 
and approach the BBN epoch with different temperatures.  
Namely, the BBN bound $\DN < 1$ implies that  
$T'/T < 0.64$. 

In this paper we study the comparative time history 
of two sectors in the early universe.    
We show that due to the temperature difference, in the 
mirror sector all key epochs as are the baryogenesis, 
nucleosynthesis, etc. proceed at somewhat different 
conditions than in the observable universe. 
The paper is organized as follows. 
In sect. 2 we describe an inflationary model 
which could provide different reheating temperatures 
between two sectors. 
The sect. 3 we show that the baryon asymmetry in the 
M-world generically should be larger than that in 
the O-sector either in the GUT or electroweak 
baryogenesis scenarios. 
Moreover, it is pretty plausible that M-baryons provide 
a significant fraction of the dark matter of the universe, 
i.e. $\Omega'_B > \Omega_B$.   
In sect. 4 we study the chemical composition of the M-sector  
and find that it should be dominantly a helium world -- 
the primordial abundance of the mirror $^4$He  
should be $2-3$ times larger than of the ordinary one. 
In sect. 5  we briefly address the problem of 
the cosmological structure formation in the presence 
of M-baryons as a dominant dark matter component.
Namely, M-baryons being a sort of self-interacting 
dark matter could provide 
interesting signatures on the CMB anisotropy,  
the large scale structure of the universe, 
the form of the galactic halos, microlensing, etc. 
Finally, in sect. 6 we summarize our findings.

\section{\bf  Inflation and post-inflation }

An attractive realization of the inflationary paradigm 
is provided by supersymmetric models of hybrid inflation. 
The symplest model is based on the 
superpotential $W=\lambda S(\Phi^2 -\mu^2)$ containing  
the inflaton field $S$ and the additional `orthogonal' 
field $\Phi$, where $\lambda$ is order 1 coupling constant 
and $\mu$ is a dimensional parameter of the order of 
the GUT scale \cite{Gia}. 
The supersymmetric vacuum is located at 
$S=0$, $\Phi = \mu$, while for the field values $\Phi=0$, 
$S > \mu$ the tree level potential has a flat valley 
with an energy density $V = \lambda^2 \mu^4$. 
Since the supersymmetry is broken by the non-vanishing 
$F$-term, $F_S=\lambda\mu^2$,    
the flat direction is lifted by radiative corrections 
and the potential of $S$ gets a slope which 
is appropriate for the slow roll conditions. 
The COBE results on the CMB anisotropy imply that 
$V^{1/4} \simeq \eps^{1/4} \times 7\cdot 10^{16}$ GeV, 
where $\eps\ll 1$ is a slow-roll parameter. 


The genetic problem of the hybrid inflation models 
concerns the choice of the initial conditions \cite{Nikos,BCT}. 
Namely, at the end of the Planck era the scalar $S$ 
should have an initial value $\sim M_P$ while $\Phi$ 
must be zero with the high accuracy over a region much 
larger than the initial horizon size $\sim M_P^{-1}$. 
In other words, the initial field configuration should  
be located just on the bottom of the $S$ valley, 
and so the energy density to start with should be 
$\sim \mu^4$, at least eight orders of magnitude smaller than 
the natural energy density $M_P^4$ at the Planck era. 
If the $\Phi$ field would have initial amplitude 
$\sim M_P$, then its oscillation should be damped 
within typical time $\sim M_P^{-1}$, otherwise 
it induces too a big curvature for the inflaton $S$ 
and violate the slow-roll conditions, thus 
preventing the onset of the inflation. 

Such a {\it Fine Tuning} can be avoided by a possible 
preheating before the onset of the inflation:  
the oscillating field $\Phi$ could decay into some 
light particles \cite{BCT}.  
In this view, one can consider a model based on 
superpotential \cite{BCT}: 
\be{W}
W_{\rm preheat} 
=\lambda S(\Phi^2 -\mu^2) + g\Phi(\Psi^2 + \Psi'^2) 
\ee 
Now all fields can have initial values order $M_P$, 
and thus the initial energy density can be $\sim M_P^4$.  
At first instants, due to oscillation of $\Phi$,  
the system behaves as matter dominated universe. 
Fast damping of $\Phi$ is a pretext of inflationary 
stage which allows the inflaton energy density $\sim \mu^4$ 
to dominate and then $S$ can slowly roll to the origin.
This function is carried by the second term in (\ref{W}) 
-- the oscillating orthogonal field $\Phi$ 
fastly decays into $\Psi$ and $\Psi'$ particles 
which have practically no contact to inflaton $S$. 
In addition, with vanishing $\Phi$, also  
effective mass terms of $\Psi$'s disappear  
and the latter fields start to behave as massless --   
they stop oscillating and freeze. 
In general, oscillations $\Psi$ and $\Psi'$  freeze 
at different amplitudes, typically $\sim \mu$, 
at which they are catched by the moment when their 
mass $g\Phi$ drops below the Hubble parameter 
(see Fig. 3 of ref. \cite{BCT}). 
When slow roll ends up, all fields start to 
oscillate around their vacuum values, $S=0$, 
$\Phi=\mu$, $\Psi,\Psi'=0$, and reheat the universe. 

Let us assume now that reheating occurs due 
to superpotential terms:  
\be{reheating}
W_{\rm reheat}= h\Psi \phi_1\phi_2 + h\Psi' \phi'_1\phi'_2
\ee
where $\phi_{1,2}$ and $\phi'_{1,2}$ are the Higgs 
doublet superfields respectively for the O- and M-sectors.  
Then different magnitudes of $\Psi$ and $\Psi'$ 
at the end of slow-roll phase should reflect into 
difference of reheating temperatures $T_R$ and $T'_R$ 
in two systems, simply because of the energy difference 
stored into $\Psi$ and $\Psi'$ oscillations which decay 
respectively into O- and M-Higgses. 
More detailed analysis of this mechanism will be presented 
elsewhere.\footnote{For other scenarios of 
the asymmetric reheating, see \cite{BDM,Venya,KST}. 
} 

We have also to make sure that after reheating 
two sectors do not come into the thermal equilibrium to 
each other, or in other words, that interactions between 
O- and M-particles are properly suppressed. 
In a supersymmetric theory this condition is fulfilled 
in a rather natural manner. In particular, the mixed  
terms of the lowest dimension in the superpotential are: 
\be{Planck} 
\frac{\beta_{ij}}{M} (l_i\phi)(l'_{j} \phi') + 
\frac{\beta}{M} (\phi_1 \phi_2)(\phi'_1 \phi'_2) ,    
\ee
which are suppressed by a large mass factor $M\sim M_P$ 
or so, and thus are safe.
The same holds true for soft supersymmetry breaking 
$F$- and $D$-terms like 
$[\frac{z}{M} (\phi_1 \phi_2)(\phi'_1 \phi'_2)]_F$, etc.,     
where $z=m_S\theta^2$ is the supersymmetry breaking spurion  
with $m_S \sim 1$ TeV. As for the
the kynetic mixing term $F^{\mu\nu} F'_{\mu\nu}$ 
between the field-strength tensors of the gauge 
factors $U(1)$ and $U(1)'$, it can be forbidden  
by embedding $G_{SM}\times G'_{SM}$ 
in the grand unified group like $SU(5)\times SU(5)'$.       


Once the O- and M-systems are decoupled already 
after reheating, at later times $t$ they will have different
temperatures $T(t)$ and $T'(t)$, and so  
different energy and entropy densities:   
\be{rho}
\rho(t) = {\pi^{2}\over 30} g_\ast(T) T^{4}, ~~~ 
\rho'(t) = {\pi^{2}\over 30} g'_\ast(T') T'^{4} ~,  
\ee
\be{s}
s(t) = {2\pi^{2}\over 45} g_{s}(T) T^{3} , ~~~ 
s'(t) = {2\pi^{2}\over 45} g'_{s}(T') T'^{3} ~.      
\ee
The factors $g_{\ast}$, $g_{s}$ and $g'_{\ast}$, $g'_{s}$ 
accounting for the effective number of the degrees of freedom 
in two systems can in general be different from each other.   
During the universe expansion, the two sectors evolve 
with separately conserved entropies. Therefore,   
the ratio $x\equiv (s'/s)^{1/3}$ is time 
independent,\footnote{ We assume 
that expansion goes adiabatically in both sectors and 
neglect the additional entropy production due to the possible 
weakly first order electroweak or QCD phase transitions.} 
while the ratio of the temperatures
in two sectors is simply given by:
\be{t-ratio}
\frac{T'(t)}{T(t)} = x \cdot 
\left[\frac{g_{s}(T)}{g'_{s}(T')} \right] ^{1/3} ~.
\ee

The Hubble expansion rate is determined by the total 
energy density $\rhb=\rho+\rho'$, $H=\sqrt{(8\pi/3) G_N\rhb}$. 
Therefore, at a given time $t$ in a radiation dominated epoch 
we have 
\be{Hubble}
H(t) = {1\over 2t} = 1.66 \sqrt{\bg(T)} \frac{T^2}{M_{Pl}} = 
1.66 \sqrt{\bg'(T')} \frac{T'^2}{M_{Pl}} ~  
\ee
in terms of O- and M-temperatures $T(t)$ and $T'(t)$, where 
\beqn{g-ast} 
\bg(T) = g_\ast (T) (1 + ax^4), ~~~~ 
\bg'(T') = g'_\ast (T')\left(1 + \frac{1}{ax^4}\right) .  
\eeqn
Here the factor $a(T,T') = [g'_\ast (T')/g_\ast (T)] 
\cdot [g_{s}(T)/g'_{s}(T')]^{4/3}$  
takes into account that for $T'\neq T$ the relativistic 
particle contents of the two worlds can be different. 
However, except for very small values of $x$, 
we have $a \sim 1$. So hereafter we always take 
$\bg(T) = g_\ast (T) (1 + x^4)$ and 
$\bg'(T') = g'_\ast (T')(1 + x^{-4})$. 
In particular, 
in the modern universe we have $a(T_0,T'_0) =1$,  
$g_s(T_0) = g'_s(T'_0)=3.91$, and $x= T'_0/T_0$, 
where $T_0,T'_0$ are the present 
temperatures of the O- and M- relic photons.\footnote{
The frozen ratio of the neutrino and photon temperatures 
in the M-sector $r'_0=T'_{\nu 0}/T'_0$ 
depends on the $\nu'$ decoupling temperature from the 
mirror plasma  which scales approximatively as
$T'_D \sim x^{-2/3} T_D$,  where $T_D= 2-3$ MeV is 
the decoupling temperature of the usual neutrinos.
Therefore, unless $x < 10^{-3}$, 
$r'_0$ has a standard value 
$r_0= T_{\nu 0}/T_0=(4/11)^{1/3}$.  
For $x < 10^{-3}$, $T'_D$ becomes larger than 
the QCD scale $\Lambda \simeq 200$ MeV,  
so that due to the mirror gluons and light quarks $u',d',s'$ 
contribution we would obtain $r'_0=(4/53)^{1/3}$, 
$g'_s(T'_0)=2.39$ and so $T'_0/T_0 \simeq 1.2 x$.  
However, in the following such small values of $x$ 
are not of our interest.} 
In fact, $x$ is the only free parameter in our model  
and it is constrained by the BBN bounds. 

The observed abundances of light elements are in 
good agreement with the standard nucleosynthesis predictions,   
when at $T\sim 1$ MeV we have $g_\ast=10.75$     
as it is saturated by photons $\gamma$, electrons $e$ 
and three neutrino species $\nu_{e,\mu,\tau}$. 
The contribution of mirror particles 
($\gamma'$, $e'$ and $\nu'_{e,\mu,\tau}$) 
would change it to $\bg =g_\ast (1 + x^4)$. 
Deviations from $g_\ast=10.75$ are usually 
parametrized in terms of the effective number 
of extra neutrino species, 
$\Delta g= \bar{g}_\ast -10.75=1.75\Delta N_\nu$.  
Thus we have:  
\be{BBN}
\Delta N_\nu = 6.14\cdot x^4 ~. 
\ee
In view of the present observational situation, 
a reliable bound is $\Delta N_\nu < 1$ \cite{io},  
which translates as $x < 0.64$. 
This limit very weakly depends on $\DN$. 
E.g. $\Delta N_{\nu} < 1.5$ implies $x < 0.70$.  

As far as $x^4\ll 1$, in a relativistic epoch 
the Hubble expansion rate (\ref{Hubble}) is dominated 
by the O-matter density and the presence of M-sector 
practically does not affect the standard cosmology 
of the early ordinary universe. 
However, even if the two sectors have the 
same microphysics, the cosmology of the early 
mirror world can be very different from the 
standard one as far as the crucial epochs like 
baryogenesis, nuclesosynthesis, etc. are concerned. 
Any of these epochs is related to an instant when 
the rate of the relevant particle process  
$\Ga(T)$, which is generically a function 
of the temperature, becomes equal to the Hubble 
expansion rate $H(T)$.
Obviously, in the M-sector these events take place  
earlier than in the O-sector, 
and as a rule, the relevant processes in the former 
freeze out at larger temperatures than in the latter. 

In the matter domination epoch the situation becomes 
different. 
In particular, we know that ordinary baryons can provide 
only a minor fraction of the present cosmological density, 
$\Omega_B = 0.01-0.06$, whereas the observational data 
indicate the presence of dark matter which  
can amount for $\Omega_{m} \sim 0.2-1$.   
It is interesting to question whether the missing 
matter density of the universe could be due to 
mirror baryons? In the next section we show that 
this situation can emerge in a pretty natural manner.

\section{Baryogenesis and mirror baryon density}

It is well known that a non-zero baryon asymmetry (BA) 
can be produced in the initially baryon symmetric universe 
if three following conditions are fulfilled: 
B-violation, CP violation and departure from the 
thermal equilibrium. 
Generally speaking, the baryogenesis scenarios 
can be divided in two categories in which 
the out of equilibrium conditions are provided 
(a) by the universe expansion itself, or (b) 
by fast phase transition and bubble nucleation.  
In particular, the latter concerns the electroweak 
baryogenesis schemes, while the former is typical 
for a GUT type baryogenesis or leptogenesis.

At present it is not possible to say which of the known 
mechanisms is responsible for the observed BA.  
We only know that the baryon to photon number density ratio 
$\eta= n_B/n_\gamma$ is restricted by the BBN constraints 
to the range $\eta = (2 - 6)\times 10^{-10}$.  
It is most likely that the BA in the M-world  
$\eta'=n'_B/n'_\gamma$ is produced by same mechanism 
and moreover, the rates of the $B$ and CP violation 
processes are parametrically the same in both cases.  
However, the out of equilibrium conditions should 
be different since at relevant temperatures the 
universe expansion is faster for the M-sector.  
Below we show that by this reason $\eta'$ typically 
emerges  larger than $\eta$ 
for either type (a) or (b) scenarios.  

The M-baryons can be of the cosmological relevance if  
$\Omega'_B$ exceeds 
$\Omega_B = 3.67\times 10^7\eta h^{-2} =0.01-0.06$,       
whereas $\Omega'_B > 1$ would overclose the universe. 
So we are interested in a situation when the ratio 
$\beta =\Omega'_B/\Omega_B$ falls in the range from 1 to 
about 100. 
Since $n'_\gamma = x^3 n_\gamma$, 
we obtain $\beta = x^3\eta'/\eta$.  
Therefore, $\eta' > \eta$ does not apriori mean that 
$\beta > 1$, and in fact there is a lower limit 
$x>10^{-2}$ or so for the relevant parameter space. 
Indeed, it arises from $x^3 = \beta \eta/\eta'$ 
by recalling that $\eta \sim 10^{-9}$, while $\eta'$ 
can be taken at most $\sim 10^{-3}$, the biggest value  
which can be principally realized in any baryogenesis 
scheme under the realistic assumptions.

\subsection{GUT Baryogenesis }

The GUT baryogenesis mechanism typically based on 
a superheavy boson $X$ undergoing the B- and CP-violating
decays into quarks and leptons. 
The following reaction rates are of relevance:
\par\noindent 
{\it Decay}:   
$\Gamma_D \sim \alpha_X M_X$ for $T\lesssim M_X$ or
$\Gamma_D \sim \alpha_X M_X^2/T$ for $T\greatsim M_X$, 
where $\alpha_X$ is the coupling strength of $X$ to fermions 
and $M_X$ is its mass; 
\par\noindent
{\it Inverse decay}: 
$\Gamma_{ID} \sim  \Gamma_D 
(M_X/T)^{3/2} \exp(-M_X/T)$ for $T\lesssim M_X$ or
$\Gamma_{ID} \sim \Gamma_D$ for $T\greatsim M_X$;
\par\noindent
{\it The $X$ boson mediated $2\leftrightarrow 2$ processes}:  
$\Gamma_S \sim n_X \sigma\sim  A \alpha_X^2T^5/(M_X^2+T^2)^{2}$, 
where the factor $A$ amounts for the possible reaction channels.

The final BA depends on a temperature  
at which $X$ bosons go out from equilibrium. 
One can introduce a parameter which measures the 
effectiveness of the decay at the 
epoch $T\sim M_X$ \cite{kolb}:    
$k=(\Gamma_D/2H)_{T=M_X}=
0.3\bg^{-1/2}(\alpha_X M_{Pl}/M_X)$.  
The larger is $k$ the longer equilibrium is 
maintained and the freeze-out abundance of $X$ boson 
becomes smaller. Hence, the resulting 
baryon number to entropy ratio, $B=n_B/s\simeq 0.14\eta$ 
is a decreasing function of $k$.    
It is approximately given as 
$B \simeq \frac{\epsilon}{g_s} F(k,k_c)$, 
where $\epsilon$ is the CP violating factor and 
\begin{equation}\label{Fkkc}
F(k,k_c)=\cases{1 & if $ k < 1$ \cr
             0.3 k^{-1}(\log k)^{-0.6} 
             & if $ 1< k <  k_c$ \cr
                 \sqrt{A\,\alpha_X\,k}\;
e^{-\frac{4}{3}(A\,\alpha_X\,k)^{1/4}} & if $ k > k_c$\cr}
\end{equation}
Here $k_c$ is a critical value defined by equation 
$k_c(\log k_c)^{-2.4}=300/(A \alpha_X)$. 
It distinguishes between the regimes $k < k_c$, 
in which inverse decay is relevant 
and $k > k_c$, in which instead $2\leftrightarrow2$ 
processes are the dominant reason for baryon damping.   

In a general context, without referring to a particular 
model, it is difficult to decide which range of parameters 
$k$ and $k_c$  can be relevant for baryogenesis. 
One can impose only the most reasonable constraints  
$g_s(T=M_X) \geq 100$ and $\epsilon \leq 10^{-2}$, and 
thus $\epsilon/g_s < 10^{-4}$ or so.
For a given mechanism responsible for the observed 
baryon asymmetry $B \sim 10^{-10}$,  
this translates into a lower bound 
$F(k,k_c) > 10^{-6}$.

\begin{figure}

\parbox{7.3cm}{
\scalebox{0.45}{
\includegraphics*[80,445][440,700]{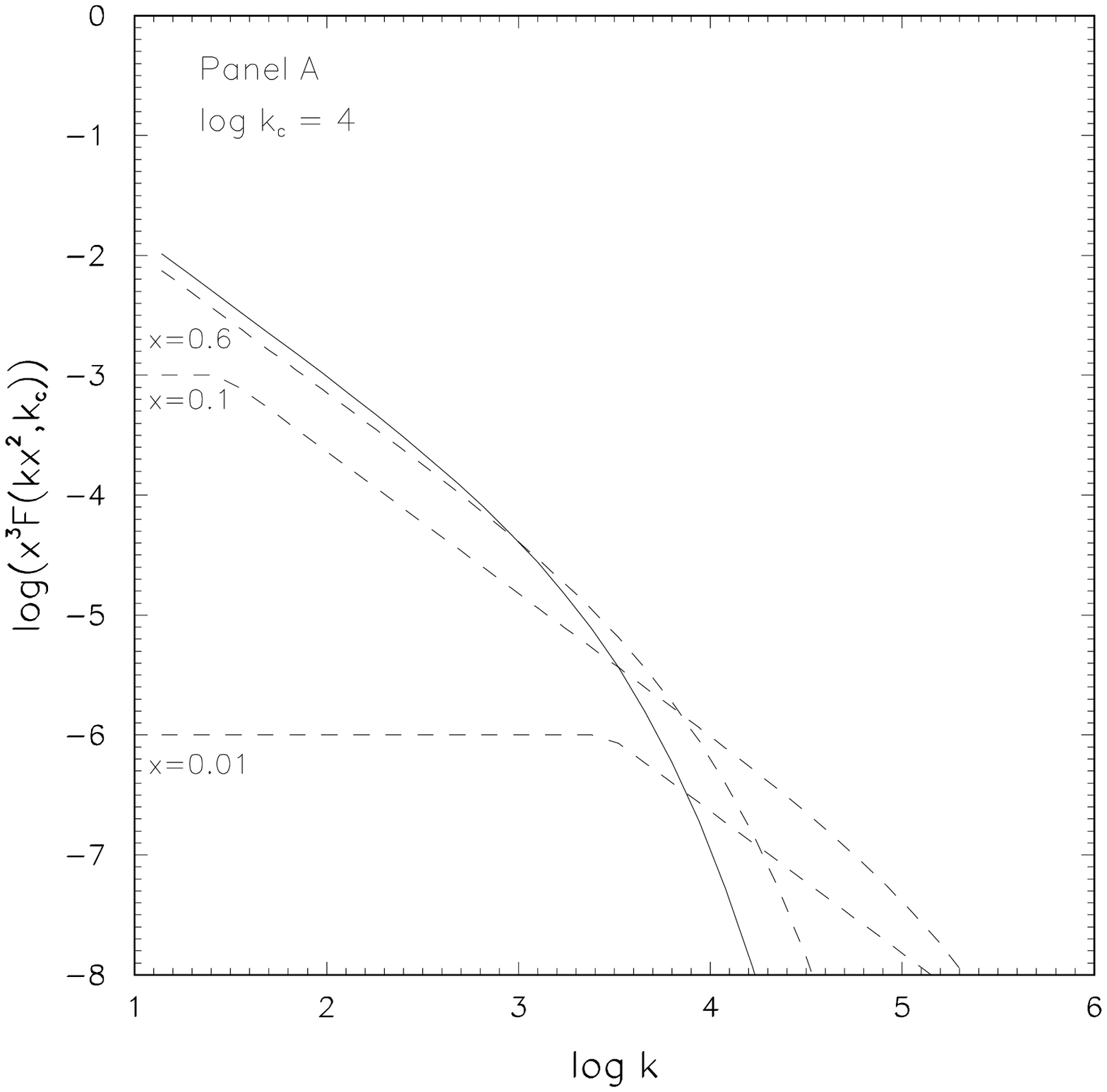}}
}
\nolinebreak
\parbox{7.3cm}{
\scalebox{0.45}{
\includegraphics*[40,445][410,700]{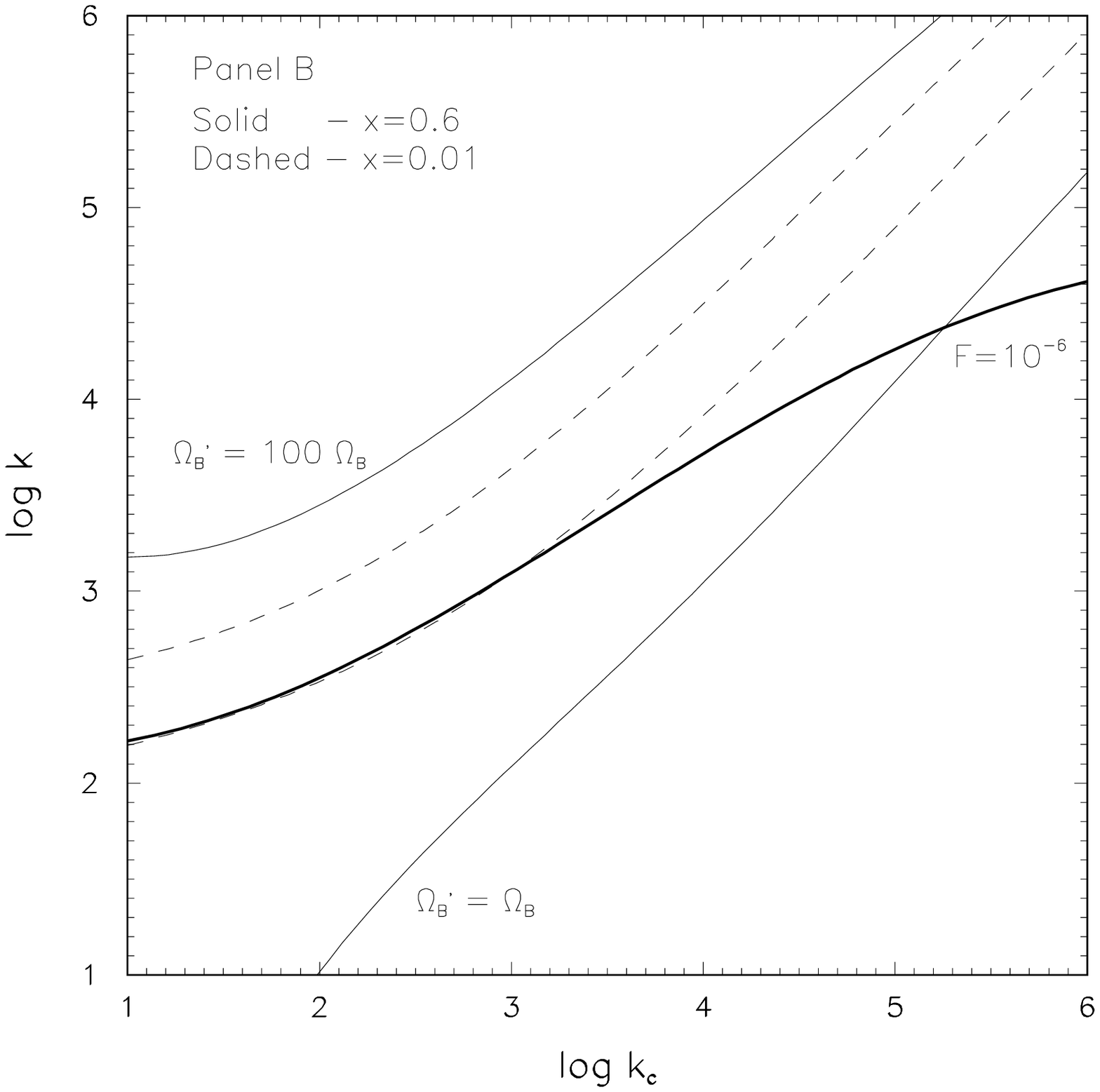}}
}

\begin{center}
\vspace{3cm}
\caption{\small {\it Panel A}. 
The combination $x^3F(kx^2,k_c)$ as a function of $k$ 
for $k_c=10^4$ and $x=0.6,\;0.1,\;0.01$ (dash).  
The solid curve corresponding to $x=1$ 
in fact measures the possible BA in the ordinary world, 
$F(k,k_c)=(g_s/\epsilon)B(k)$.  
{\it Panel B}. The curves confining the parameter region 
in which $\beta=\Omega'_B/\Omega_B$ varies from 1 to 100, 
for $x=0.6$ (solid) and for $x=0.01$ (dash). 
The parameter area above thick solid curve  
corresponds to $F(k,k_c) < 10^{-6}$ and it  
is excluded by the observable value of $\eta$. } 
\end{center}
\end{figure}

The presence of the mirror sector practically 
does not alter the ordinary baryogenesis.  
The effective particle number is 
$\bar g_\ast (T) = g_\ast(T)(1+x^4)$ and thus   
the contribution of M-particles to the Hubble constant 
at $T\sim M_X$ is suppressed by a factor $x^4$.

In the mirror sector everything should occur 
in a similar way, apart from the fact that now 
at $T'\sim M_X$ the Hubble constant is not 
dominated by the mirror species but by ordinary ones:   
$\bar g'_\ast (T')\simeq g'_\ast (T')(1+ x^{-4})$.  
As a consequence, we have 
$k' = (\Gamma'_D/2H)_{|T'=M_X} = k x^2$. 
Since the value of $k_c$ is the same in the two sectors,
the mirror baryon asymmetry can be
simply obtained by replacing 
$k\rightarrow k'=kx^2$ in eq.\ref{Fkkc},
i.e. $B'=n'_B/s' \simeq (\epsilon/g'_s) F(k'=kx^2,k_c)$.
Since $F$ is a decreasing function of $k$, then 
for $x < 1$ we have $F(kx^2,k_c) > F(k,k_c)$ and  
thus we conclude that the mirror world always gets a  
{\it larger} BA than the visible one, $B' > B$. 

However, this does not apriori mean 
that $\Omega'_B$ is always larger than $\Omega_B$. 
Since the entropy densities are related as $s'/s=x^3$,   
for the ratio $\beta =\Omega'_B/\Omega_B$ we have:
\be{B-ratio}
\beta (x) =  \frac{ n'_B}{n_B} = \frac{B's'}{Bs} =  
x^3\;\frac{F(kx^2,k_c)}{F(k,k_c)} ~.
\ee
The behaviour of the factor $x^3 F(kx^2,k_c)$ 
as a function of $k$ for different values of the 
parameter $x$ is given in the Fig. 1A. 
Clearly, in order to have $\Omega'_B > \Omega_B$ 
the function $F(k,k_c)$ have to decrease
faster than $k^{-3/2}$ between $k'=kx^2 $ and $k$.
Closer inspection of the function \ref{Fkkc} reveals  
that the M-baryons can be overproduced only if 
$k$ is order $k_c$ or larger.  In other words, 
the relevant interactions in the observable sector 
maintain equilibrium longer than in the mirror one, 
and thus ordinary BA can be suppressed by an 
exponential Boltzmann factor while the mirror BA 
could be produced still in non-exponential regime 
$k' < k_c$.  

In Fig. 1B we show the parameter region in which
$\beta =\Omega'_B/\Omega_B$ falls in the range $1-100$,  
in confront to the parameter area 
excluded by condition $F(k,k_c)>10^{-6}$.  
We see that for $x=0.6$ there is an allowed parameter 
space in which $\beta$ can reach values up to 10, 
but $\beta=100$ is excluded. 
For a limiting case $x=10^{-2}$, as it was expected, 
the parameter space for $\beta > 1$ becomes 
incompatible with $F(k,k_c)> 10^{-6}$. 
For intermediate values of $x$, say $x\sim 0.1-0.3$, also 
the values $\beta \sim 100$ can be compatible.  
 
The above considerations can be applied also in the 
context of leptogenesis. One should remark, however, 
that potentially both the GUT baryogenesis or 
leptogenesis scenarios are in conflict with   
the supersymmetric inflation scenarios, 
because of the upper limit on the reheating 
temperatures about $T_R < 10^9$ GeV from the thermal 
production of gravitinos \cite{TR-therm}.  
Moreover, it was shown recently that the non-thermal 
gravitino production can impose much stronger limits, 
$T_R < 10^5$ GeV or so \cite{TR-nontherm}. 
This problem can be fully avoided in the 
electroweak baryogenesis scenario, which instead 
is definitely favoured by the supersymmetry.

\subsection{Electroweak Baryogenesis }

The electroweak (EW) baryogenesis mechanism is based 
on the anomalous B-violating processes induced by the 
sphalerons which are quit rapid at high temperatures,  
but become much slower when temperature drops 
below 100 GeV. 
A succesfull scenario needs the first order EW phase 
transition and sufficient amount of CP violation,  
which conditions can be satisfied in the frames  
of the supersymmetric standard model, for certain 
parameter ranges \cite{bau}. 

The characteristic temperature scales of the 
electroweak phase transition are fixed entirely by the 
form of the finite temperature effective potential:  
\be{effective}
V(\phi,T)=D(T^2-T_0^2)\phi^2- E T\phi^3 + \lambda_T\phi^4 
\ee 
where all parameters can be expressed in terms of 
the fundamental couplings in the Lagrangian.  
For large temperatures, $T \gg 100$ GeV, the electroweak  
symmetry is restored and $V(\phi,T)$ has a minimum 
at $\phi=0$. 
With the universe expansion the temperature drops,  
approaching the specific values which define 
the sequence of the phase transition. 
These are all in the 100 GeV range and ordered as  
$T_1> T_c > T_b > T_0$. 
Namely, below $T=T_{1}$ the potential gets 
a second local minimum $\phi_{+}(T)$. 
At the critical temperature $T=T_c$ the latter becomes 
degenerate with the false vacuum $\phi=0$. 
At temperatures $T<T_c$ to the true vacuum 
state $\phi=\phi_+(T)$ becomes energetically favoured, 
and transition to this state can occur via thermal 
quantum tunneling, through the nucleation of the bubbles 
which then expand fastly, percolate and finally fill the 
whole space within a horizon by the true vacuum.  

The bubble production starts 
when the free energy barrier separating the two minima 
becomes small enough. 
The bubble nucleation temperature $T_b$ is defined as 
a temperature at which the probability for a single
bubble to be nucleated within a horizon volume becomes 
order one: 
\be{nucl}
P(T_b)= \omega \left( \frac{T_c}{H(T_c)} \right)^4 
\left(1 -\frac{T_b}{T_c}\right)
e^{-\frac{F_c(T_b)}{T_b}}\sim 1 
\ee
where $F_c(T)$ is the free energy and $\omega$ is an 
order one coefficient \cite{bau}. 
In particular, in the limit of thin wall approximation 
we have:
\be{thin}
\frac{F_c(T)}{T}= \frac{64 \pi}{81} \frac{E}{(2 \lambda)^{3/2}}
\left(\frac{T_c-T_0}{T_c-T} \right)^2 . 
\ee
The condition (\ref{nucl}) results into large values 
of $F_c(T_b)/T_b$, typically order $10^2$. 


Once the bubble nucleation and expansion rate is larger 
than the Hubble parameter, 
the out of equilibrium condition for anomalous 
B-violating processes is provided by 
the fast phase transition itself.  
The BA can be produced inside the bubbles due to the 
CP violation since the quarks and antiquarks have different 
reflection coefficients on the bubble wall. 
The baryogenesis rate is completely 
independent from the universe expansion 
and it occurs practically in one instant as compared 
to the cosmological time scale of this epoch.

As for the mirror sector, it is described by the same 
Lagrangian as the ordinary one and so the effective 
thermal potential of the mirror Higgs $V(\phi',T')$ 
has the same form as (\ref{effective}). 
Then the temperature scales which are defined 
entirely by the form of the effective potential 
should be exactly the same for O- and M-sectors.  
Namely, $T'_1=T_1$ and $T'_c=T_c$.

The equation (\ref{nucl}) is the same for the M-sector 
apart of the fact that the corresponding 
Hubble constant is different: 
$H(T'=T_c)= x^{-2} H(T=T_c)$. Therefore, we obtain:  
\beq
\frac{F_c(T'_b)}{T'_b}=\frac{F_c(T_b)}{T_b} + 8 \log{x}   
\ee 
which in turn tells that the bubble nucleation temperatures 
in two sectors are practically equal: 
$T'_{b}=T_{b}(1+0.01 \log x)$. 
(Clearly, the phase transition in M-sector occurs at earlier  
time than in O-sector: $t'_b \simeq x^{-2} t_b$.)
The reason is that between the temperature
scales $T_c$ and $T_0$ the free energy $F_c(T)$ is a 
rapidly changing function but the change 
in temperature itself is insignificant.
Hence, we expect that the initial BA's produced right 
at the phase transition to be the same in O- and M-sectors: 
$B(T=T_b)=B'(T'=T_b)$.

However, the instantly produced baryon number can be still 
washed out by the sphaleron interactions. 
The anomalous B-violation rate 
$\Gamma(T) \sim \exp[-F(T)/T]$, where $F(T)$ 
is the sphaleron free energy at finite $T$,  
may be large enough inside the bubble as 
far as the temperature is large. But it quickly falls 
as the temperature decreases, and baryon number 
freezes out as soon as $\Gamma(t)$ drops below $H(t)$ 
(\ref{Hubble}).   
The wash-out equation $dB/dt=-\Gamma(t) B$ can be 
rewritten as 
\be{wah-out}
\frac{dB}{B}= 
\frac{\Gamma(T)}{H T}dT
\ee
and integrated. Then the final BA in the O- and M-sectors can 
be expressed respectively as 
\be{B-EW}
B=B(T_b) D^{(1+x^4)^{-1/2}}, ~~~~~~
B'=B(T_b) D^{x^2(1+x^4)^{-1/2}} , 
\ee
where $D<1$ is the baryon number depletion factor:   
\be{D}
D=\exp\left[- 0.6 g_{\ast}^{-1/2} M_{Pl} 
\int_{0}^{T_{b}} dT \frac{\Gamma(T)}{T^3} \right] ~, 
\ee
and  $g_{\ast}\sim O(100)$ in the supersymmetric standard
model. Thus we always have $B' > B$, while 
the M-baryon mass density relative to O-baryons reads:
\be{baew}
\beta (x) = \frac{\Omega'_B}{\Omega_B} = x^3 \frac{B'}{B}
=x^3 D^{-K(x)}, 
  ~~~~~ K(x)= \frac{1-x^2}{\sqrt{1+x^4} } ~. 
\ee

\begin{figure}[t]
\begin{center}
\epsfig{file=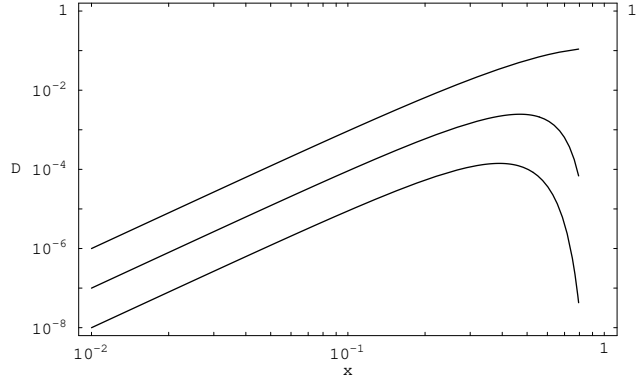,height=5cm}
\end{center}
\caption{\small The contours of 
$\beta =\Omega_B'/\Omega_B$ in the plane of the 
parameters $x$ and $D$, corresponding to 
$\beta =1,10$ and 100 from top to bottom. 
}
\label{fig2}
\end{figure}

Lacking a precise theory for non-perturbative sphaleron 
transitions in the broken phase, 
the exact value of $D$ one cannot be calculated  
even in the context of concrete models. 
If $D \sim 1$, the wash-out is ineffective 
and practically all BA produced right at the bubble 
nucleation is conserved. In this case $\Omega'_B$ 
should smaller than $\Omega_B$. 
However, if $D$ is enough small, one can achieve 
sufficiently large $\Omega'_B$.  
The contourplot for the  parameters $x$ and $D$ 
for which $\beta$ falls in the range $1-100$ 
is given in Fig. 2. 
For small $x$ we have essentially 
$\beta \simeq x^3 D^{-1}$ and thus  
$100 > \beta > 1$ requires a depletion factor 
in the interval $D = (10^{-2} -1) x^3$. 
Once again, for $x \sim 10^{-2}$ one needs  
the marginal values $D\sim 10^{-8} - 10^{-6}$
below which the observable BA $B\sim 10^{-10}$ 
cannot be produced at all.

\section{Primordial Nucleosynthesis and mirror helium abundance}

The time scales relevant for standard BBN 
are defined by the ``freeze-out'' temperature
of weak interactions $T_{W}\simeq 0.8$ MeV ($t_W\sim 1$ s) 
and by the ``deuterium bottleneck'' temperature 
$T_{N}\simeq 0.07$ MeV ($t_N \sim 200$ s) \cite{kolb}. 
When $T > T_{W}$, weak interactions transform neutrons 
into protons and viceversa and keep them in chemical 
equilibrium. The neutron abundance $X_n = n_n/n_B$, 
defined as the ratio of neutron to baryon  densities, 
is given by $X_{n}(T)=[1+\exp(\Delta m/T)]^{-1}$,
%
where $\Delta m\simeq 1.29$ MeV is the neutron-proton 
mass difference.
For $T < T_W$ the weak reaction rate 
$\Gamma_W \simeq G_F^2 T^5$ 
drops below the Hubble expansion rate 
$H(T) \simeq 5.5 T^2/M_{Pl}$, 
the neutron abundance freezes out 
at the equilibrium value $X_{n}(T_{W})$ 
and it then evolves only due to the neutron decay:
$X_{n}(t)=X_{n}(T_{W})\exp(-t/\tau)$, 
where $\tau=886.7$ s is the neutron lifetime. 

At temperatures $T > T_{N}$, the process 
$p+n \leftrightarrow d+\gamma$ is faster 
than the universe expansion, and 
free nucleons and deuterium are in chemical equilibrium.  
The light element nucleosynthesis essentially begins 
when the system cools down to the temperature 
\be{T_N}
T_{N}\simeq\frac{B_{d}}{-\ln(\eta)+1.5\ln(m_{N}/T_N)} 
\simeq 0.07 ~ {\rm MeV}, 
\ee
where $B_{d}=2.22$ MeV is the deuterium binding energy,    
and $m_{N}$ is the nucleon mass. 
Below this temperature the deuterium abundance starts to 
grow which in turn allows to produce also the heavier nuclei. 
Nearly all neutrons present at this time are finally captured 
in $^{4}$He nuclei, due to the large binding energy of 
the latter. Thus, the primordial 
$^{4}$He mass fraction is:
\be{helium}
Y_{4}\simeq2X_{n}(t_{N})=
\frac{2\exp(-t_N/\tau)}{1+\exp(\Delta m/T_{W})} \simeq 0.24.
\ee

As we have already discussed, the presence of the mirror 
sector with a temperature $T'\ll T$ has practically 
no impact the standard BBN. In fact, the limit $x < 0.64$ 
has been set by uncertainties of the present observational 
situation. 
In the mirror sector nucleosynthesis proceeds along the 
same lines. However, the impact of the O-world for the 
mirror BBN is dramatic. 

For any given temperature $T'$, now we have 
$H(T') \simeq 5.5 (1+x^{-2}) T'^2/M_{Pl}$ 
for the Hubble expansion rate. 
Therefore, the freeze-out temperature 
$T'_W=(1+x^{-4})^{1/6} T_W$ is larger than $T_{W}$,  
whereas the time scales as $t'_W = t_W/(1+x^{-4})^{5/6}$.
In addition, $\eta'$ is different from 
$\eta\simeq 5 \times 10^{-10}$. 
However, since $T_N$ depends on baryon density 
only logarithmically (see (\ref{T_N})), 
the temperature $T'_N$ remains essentially
the same as $T_N$, while the time $t'_N$
scales as $t'_N = t_N/(1+x^{-4})^{1/2}$. 
Thus, for the mirror $^4$He mass fraction we obtain:  
\be{m_helium}
Y'_{4}\simeq 2X'_n(t'_N)= 
\frac{ 2\exp[-t_N/\tau(1+x^{-4})^{1/2}] }
{1+\exp[\Delta m/T_W(1+x^{-4})^{1/6} ] } ~. 
\ee
We see that $Y'_{4}$ is an increasing function of $x^{-1}$.  
In particular, for $x\rightarrow 0$ 
one has $Y'_{4}\rightarrow 1$.

\begin{figure}[t]
\begin{center}
\epsfig{file=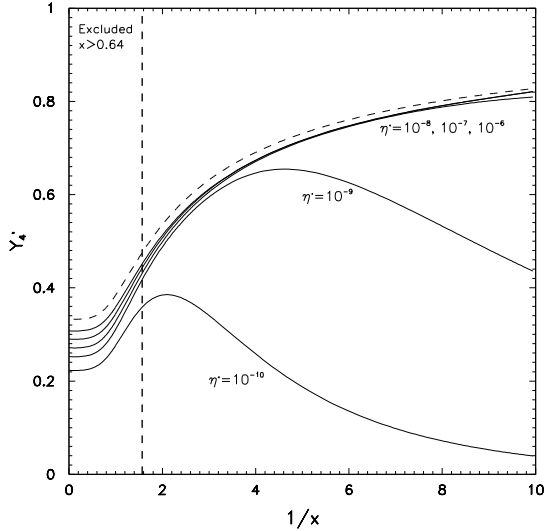,height=7cm}
\end{center}
\vspace{-0.5cm}
\caption{\small The primordial mirror $^4$He mass fraction 
as a function of $x$. The dashed curve represents the
approximate result of eq. (\ref{helium}).
The solid curves obtained via exact numerical 
calculation correspond, from bottom to top, 
to $\eta'$ varying from $10^{-10}$ to $10^{-6}$.}
\label{bbn}
\end{figure}

In reality, eq. (\ref{m_helium}) is not valid 
for small $x$, since in this case deuterium production 
through reaction $n+p\leftrightarrow d+\gamma$ can become 
ineffective.
By a simple calculation one can make sure that 
for $x < 0.3 \cdot (\eta'\times 10^{10})^{-1/2}$,
the rate at which neutrons are captured to 
form the deuterium nuclei,  
$\Gamma '_N = n'_p \sigma_N \sim \eta' n'_\gamma \sigma_N$, 
where $n'_\gamma \sim T'^3$ is the M-photons density 
and $\sigma_N \simeq 4.5\cdot 10^{-20}$ cm$^3$ s$^{-1}$ 
is the thermal averaged cross section, becomes smaller 
than the Hubble rate $H(T')$ for temperatures 
$T' > T'_N$.
 In this case M-nucleosynthesis is inhibited, because 
the neutron capture processes becomes ineffective
before deuterium abundance grows enough to initiate 
the synthesis of the heavier elements. 
Therefore, for any given $\eta'$, $Y'_4$ first increases 
with increasing $1/x$, reaches a maximum and then starts 
to decrease. The true dependence of $Y'_4$ on the $x$ 
computed for different values of $\eta'$ by standard 
BBN code \cite{Kawano}, is presented in Fig. \ref{bbn}.  
The Hubble expansion rate of the mirror world  
was implemented, for each value of $x$, by taking   
an effective number of extra neutrinos 
as $\Delta N_\nu = 6.14\cdot x^{-4}$. 

We have to remark, however, that in the most interesting 
situation when 
$\beta =\Omega'_B/\Omega_B = x^3 \eta'/\eta > 1$, 
the condition $x < 0.3 \cdot (\eta'\times 10^{10})^{-1/2}$ 
is never fulfilled and the behaviour of $Y'_4$ is well 
described by the approximate formula (\ref{m_helium}). 
Hence, in this case $Y'_4$ is always bigger than $Y_4$.   
In other words, if dark matter of the universe 
is represented by the baryons of the mirror sector,  
it should contain considerably bigger fraction of  
primordial $^4$He than the ordinary world.

\section{Mirror baryons as dark matter}

We have shown that mirror baryons could provide 
a significant contribution to the energy density
of the universe and thus they could constitute a 
relevant component of dark matter. 
Immediate question arises: 
how the mirror baryon dark matter (MBDM) behaves 
and what are the differences from the more familiar dark 
matter candidates as the cold dark matter (CDM), 
the hot dark matter (HDM) etc. 
In this section we briefly address the 
possible observational consequences of 
such a cosmological scenario.  

In the most general context, the present energy 
density contains relativistic (radiation) component 
$\Omega_r$, non-relativistic (matter) component 
$\Omega_m$ and the vacuum energy 
density $\Omega_\Lambda$ (cosmological term). 
According the inflationary paradigm the universe 
should be almost flat,  
$\Omega_0=\Omega_m + \Omega_r + \Omega_\Lambda \approx 1$,  
which well agrees with the recent results  
on the CMB anisotropy \cite{Boom}. 
The Hubble parameter is known to be 
$H_0 = 100 h$ km s$^{-1}$ Mpc$^{-1}$ with   
$h =0.6-0.8$, and for redshifts of the cosmological 
relevance, $1+z = T/T_0 \gg 1$, it becomes  
\be{H}
H(z)= H_0 \left[\Omega_{r}(1+z)^4
+ \Omega_{m} (1+z)^3 \right]^{1/2}  . 
\ee
In the context of our model, the relativistic fraction 
is represented by the ordinary and mirror 
photons and neutrinos,  
$\Omega_rh^2=4.2\times 10^{-5}(1+x^4)$, and    
contribution of the mirror species is negligible  
in view of the BBN constraint $x< 0.64$. 
As for the non-relativistic component, 
it contains the O-baryon fraction $\Omega_B$ and
the M-baryon fraction $\Omega'_B = \beta\Omega_B$,     
while the other types of dark matter, e.g. the CDM, 
could also present and so 
$\Omega_m=\Omega_B+\Omega'_B+\Omega_{\rm CDM}$.\footnote{
In the context of supersymmetry,   
the CDM component could exist in the form of 
the lightest supersymmetric particle (LSP).  
It is interesting to remark that the mass fractions  
of the ordinary and mirror LSP are related as 
$\Omega'_{\rm LSP} \simeq x\Omega_{\rm LSP}$. 
In addition, a significant HDM component $\Omega_\nu$ 
could be due to neutrinos with order eV mass. 
The contribution of the mirror neutrinos 
scales as $\Omega'_\nu = x^3 \Omega_\nu$ and thus   
it is irrelevant.  
}
At present it is not completely clear what is the matter 
fraction in the universe. Many observational data favour 
$\Omega_m \sim 0.3$ while the rest 
of the energy density is due to the cosmological term, 
$\Omega_\Lambda \sim 0.7$, but also $\Omega_m \sim 1$ 
cannot be excluded.   
%
  
The important moments for the structure formation  
are related to the matter-radiation equality (MRE) epoch 
which occurs at the redshift (we denote 
$(\Omega_{m}h^2)_{0.2} = (\Omega_{m}h^2/0.20)$):   
\be {z-eq} 
1+z_{\rm eq}= \frac{\Omega_m}{\Omega_r} \approx 
 2.4\cdot 10^4 \frac{\Omega_{m}h^2}{1+x^4} =  
 4800 \times (\Omega_{m}h^2)_{0.2} 
\ee
and to the plasma recombination and matter-radiation 
decoupling (MRD) epochs.  
The latter takes place only after the most of 
electrons and protons recombine into neutral hydrogen   
and the free electron number density  $n_{e}$ diminishes,   
so that the photon scattering rate 
$\Gamma_\gamma=n_{e}\sigma_{T}=X_{e}\eta n_{\gamma} \sigma_{T}$ 
drops below the Hubble expansion rate $H(T)$,  
where $\sigma_T=6.65\cdot 10^{-25}$ cm$^{2}$ 
is the Thomson cross section. 
In condition of chemical equilibrium, 
the fractional ionization $X_e=n_e/n_B$ is given by 
the Saha equation, which for $X_e \ll 1$ reads: 
\be{Saha} 
X_e \approx (1-Y_4)^{1/2}\; \frac{0.51}{\eta^{1/2}} 
\left(\frac{T}{m_e}\right)^{-3/4} e^{-B/2T} 
\ee 
where $B=13.6$ eV is the hydrogen binding energy.  
Thus we obtain the familiar result 
that in our universe the MRD takes place 
in the matter domination period, at the temperature 
$T_{\rm dec} \simeq 0.26$ eV which 
corresponds to redshift  
$1+z_{\rm dec}=T_{\rm dec}/T_0 \simeq 1100$. 

\begin{figure}[t]
\begin{center}
\epsfig{file=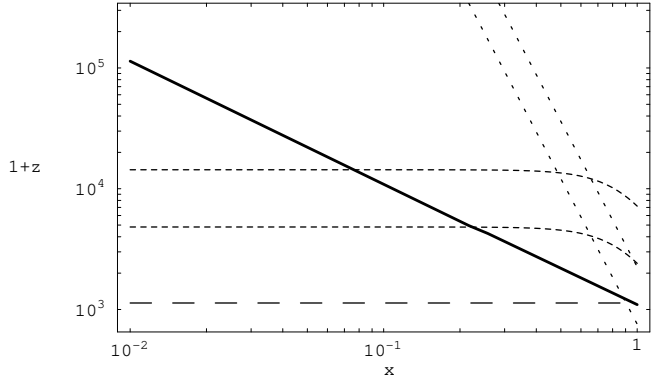,height=5cm}
\end{center}
\vspace{-0.5cm}
\caption{\small The M-photon decoupling redshift 
$1+z_{dec}'$ as a function of $x$ (solid).  
The long-dash line marks the ordinary decoupling   
redshift $1+z_{\rm dec} = 1100$. 
We also show the matter-radiation equality redshift 
$1+z_{\rm eq}$ and the mirror Jeans-horizon mass equality  
redshift $1+z'_c$, 
for the cases $\Omega_m h^2 =0.2$ (respectively lower dash 
and lower dott) and $\Omega_m h^2=0.6$ (upper dash and dott).  
}
\label{fig3}
\end{figure}

The MRD temperature in the M-sector $T'_{\rm dec}$ 
can be calculated following the same lines as in 
the ordinary one. 
Due to the fact that in either case the 
photon decoupling occurs when the exponential factor 
in eq. (\ref{Saha}) becomes very small, 
we have $T'_{\rm dec} \simeq T_{\rm dec}$, 
up to small logarithmic corrections related to 
$\eta'$, $Y'_4$ different from $\eta$, $Y_4$. Hence 
\be{z'_dec}
1+z'_{\rm dec} \simeq x^{-1} (1+z_{\rm dec}) 
\simeq 1.1\cdot 10^3 x^{-1} 
\ee
so that the MRD in the M-sector occurs earlier 
than in the ordinary one. Moreover, for $x$ less than 
$x_{\rm eq}=0.23(\Omega_m h^2)_{0.2}^{-1}$,    
the mirror photons would decouple  
yet during the radiation dominated period  
(see Fig. \ref{fig3}).

Let us discuss now cosmological evolution of the MBDM. 
The relevant length scale for the gravitational 
instabilities is characterized by the mirror Jeans scale 
$\lambda'_J \simeq v'_s (\pi/G\rho)^{1/2}$,  
where $\rho(z)$ is the matter density at a given redshift 
$z$ and $v'_s(z)$ is the sound speed in the M-plasma.   
The latter contains more baryons and less photons than the 
ordinary one, $\rho'_B=\beta \rho_B$ and 
$\rho'_\gamma = x^4\rho_\gamma$.  
Let us consider for simplicity the case  
when dark matter of the universe is entirely due to 
M-baryons, $\Omega_m\simeq\Omega'_B$. 
Then we have:  
\be{sound} 
v'_s(z) \simeq \frac{c}{\sqrt3}
\left(1+ \frac{3\rho'_B}{4\rho'_\gamma}\right)^{-1/2} 
\approx \frac{c}{\sqrt3} 
\left[ 1 +\frac34\left(1+\frac{1}{x^4}\right)
\frac{1+z_{\rm eq}}{1+z}\right]^{-1/2} . 
\ee 
Hence, for the redshifts 
of the cosmological relevance, $z\sim z_{\rm eq}$, 
we have $v'_s \sim 2x^2 c/3 \ll c/\sqrt{3}$, 
quite in contrast with the ordinary world, 
where $v_s \approx c/\sqrt{3}$ practically 
till the photon decoupling, $z=1100$.  

The M-baryon Jeans mass 
$M'_J =\frac{\pi}{6} \rho_m \lambda'^3_J$ reaches the 
maximal value at $z=z'_{\rm dec}\simeq 1100/x$, 
$M'_J(z'_{dec}) \simeq 6 \cdot 10^{17}\times 
x^6 [1+(x_{\rm eq}/x)]^{-3/2} 
(\Omega_m h^2)_{0.2}^{-2} ~ M_\odot$. 
Notice, however, that $M'_J$ becomes smaller than the 
Hubble horizon mass $M_H = \frac{\pi}{6} \rho H^{-3}$ 
starting from a redshift
$z_c= 750 x^{-4}(\Omega_m h^2)_{0.2}$, which is   
about $z_{\rm eq}$ for $x=0.64$, but 
it sharply increases for smaller values of $x$ 
(see Fig. 4).    
So, the density perturbation scales which enter 
horizon at $z \sim z_{\rm eq}$ have mass larger 
than $M'_J$ and thus undergo uninterrupted linear growth 
immediately after $t=t_{\rm eq}$. 
The smaller scales for which $M'_J > M_H$  
instead would first oscillate.  
Therefore, the large scale structure 
formation is not delayed even if the mirror MRD epoch 
did not occur yet, i.e. even if $x> x_{\rm eq}$.  
The density fluctuations start to grow in the M-matter 
and the visible baryons are involved later, after being 
recombined, when they rewritte the spectrum of already 
developed mirror structures.

The main feature of the MBDM scenario is that the 
M-baryon density fluctuations should undergo the 
strong collisional damping around the time of 
M-recombination.
The photon diffusion from the overdense to underdense
regions induce a dragging of charged particles  
and wash out the perturbations at scales smaller than the 
mirror Silk scale $\lambda'_S \simeq 
10\times f(x)(\Omega_m h^2)_{0.2}^{-3/4}$ Mpc,    
where $f(x)=x^{5/4}$ for $x > x_{\rm eq}$, 
and $f(x) = (x/x_{\rm eq})^{3/2} x_{\rm eq}^{5/4}$ 
for $x < x_{\rm eq}$. 

Thus, the density perturbation scales which can run 
the linear growth after the MRE epoch are limited by the 
length $\lambda'_S$.
This could help in avoiding the excess of small scales 
(of few Mpc) in the large scale power spectrum without 
tilting the spectral index.\footnote{One should keep 
in mind also delaying of the MRE moment, 
e.g. by taking $\Omega_m h^2 \sim 0.2$. 
This would correspond to the case when M-baryons provide 
only a fraction of the present energy density, say 
$\Omega'_B \sim 0.3$, while the rest is due to the 
cosmological term, $\Omega_\Lambda \sim 0.7$.}  
The smallest perturbations that survive the 
Silk damping will have the mass 
$M'_S \sim x^{15/4}(\Omega_m h^2)_{0.2}^{-5/4}  
10^{13}~ M_\odot $, 
which should be less than $2\times 10^{12} ~ M_\odot$ 
in view of the BBN bound $x <0.64$.  
Interestingly, for $x\sim x_{\rm eq}$ we have  
$M'_S \sim 4\times 10^{10}~ 
(\Omega_m h^2)_{0.2}^{-5} ~M_\odot$, 
a typical galaxy mass. 

To some extend, the cutoff effect is  
analogous to the free streaming damping in the case of 
warm dark matter (WDM), but there are important  
differences. The point is that alike usual baryons, 
the MBDM should show acoustic oscillations 
whith an impact on the large scale power spectrum. 
In particular, it is tempting to imagine 
that the M-baryon oscillation effects  
are related to the anomalous features  
observed at $100 h^{-1}$ Mpc clustering \cite{Einasto}.  


In addition, the MBDM oscillations transmitted 
via gravity to the ordinary baryons,  
could cause observable anomalies in the CMB 
angular power spectrum for $l$'s larger than 200. 
This effect can be observed only if the M-baryon Jeans 
scale $\lambda'_J$ is larger than the Silk scale 
of ordinary baryons, $\lambda_S \simeq 
3 \beta^{1/2} (\Omega_m h^2)_{0.2}^{-3/4}$ Mpc, 
which sets a principal cutoff for CMB oscillations 
around $l\sim 1200$. 
As we have seen above, this would require enough large 
values of $x$, near the upper bound set by 
the BBN constraints: $x \simeq 0.6$ or so. 
The detailed analysis of this effect will be given elsewhere.   
In our opinion, together with the possible effects on 
the large scale power spectrum, it can provide a most 
direct test for the MBDM and can be verified by 
the next CMB experiments with higher sensitivity.  

Clearly, for small $x$ the M-matter recombines 
before the MRE moment, and thus it should rather manifest 
as the CDM as far as the large scale structure is concerned. 
However, there still can be crucial difference at 
smaller scales which already went non-linear,  
like galaxies. In our scenario, dark matter in 
galaxies and clusters can contain the mixed CDM and 
MBDM components, and can be even constituted entirely by 
the mirror baryons.
Then one can question whether the MBDM distribution 
in halos can be different from that of the CDM? 
Namely, simulations show that the CDM forms triaxial 
halos with a density profile too clumped towards the 
center, and overproduce the small substructures within 
the halo. As for the MBDM, it constitutes a sort of 
collisional dark matter and thus potentially could avoide 
these problems, at least clearly the one related with 
the excess of small substructures. 

The halo distribution in galaxies depends on the mass $M$ 
and on the self-scattering cross section $\sigma$ 
of dark matter.  
In our case it mainly consists of the mirror hydrogen atoms,  
and so $M_H\simeq 1$ GeV and $\sigma_H\sim 10^{-16}$ cm$^2$.   
At the first glance, this is in strong discrepancy 
with the range $\sigma/M\sim 10^{-23}-10^{-24}$ cm$^2$/GeV  
preferred by the analysis of ref. \cite{SS}.\footnote{ 
In the case of asymmetric M-world with 
$\zeta=v'/v \gg 1$, the mirror electron mass should 
scale as $m'_e\simeq \zeta m_e$ while the nucleon mass 
remains $\sim 1$ GeV. Therefore, the elastic 
scattering cross-section of the M-hydrogen atoms 
scales as $\sigma'_H \sim \zeta^{-2} \sigma_H$ and so 
for $\zeta \gg 1$ one could obtain a  
reasonably small cross section.} 

However, one has to take into account the possibility 
that during the galaxy evolution 
the bulk of the M-baryons could fastly fragment 
into the stars. 
A difficult question to address here    
is related to the star formation in the M-sector, 
also taking into account that its temperature/density 
conditions and chemical contents 
are much different from the ordinary ones. 
In any case, the fast star formation would 
extint the mirror gas and thus 
could avoide the M-baryons to form disk galaxies 
as ordinary baryons do. The M-protogalaxy,    
which at certain moment before disk formation 
essentially becomes the collisionless system of the 
mirror stars,  
could maintain a typical elliptical structure.\footnote{ 
In other words, we speculate on the possibility 
that the M-baryons form mainly the elliptical galaxies.  
For a comparison, in the ordinary world 
the observed bright galaxies are mainly spiral while  
the elliptical galaxies account about $20 ~\%$ of them. 
Remarkably, the latter contain old stars, very little dust 
and show no sign of active star formation.
} 
Certainly, in this consideration also the galaxy merging 
process should be taken into account.  
As for the O-matter, within the dark M-matter halo it should 
typically show up as an observable elliptic or spiral galaxy,  
but some anomalous cases can be also possible, 
like certain types of irregular galaxies or even 
dark galaxies dominantly made out of M-baryons. 

The central part of halo can nevertheless contain a 
large amount of the ionized mirror gas and it is not 
excluded that it can have a quasi-spherical form. 
Even if the stellar formation is very efficient, 
the massive mirror stars in the dense central region 
fastly evolve\footnote{
Since  mirror matter contains more 
helium, the mirror stars should evolve much 
faster than the ordinary ones.
}
and explode as supernovae, leaving behind the compact 
objects like neutron stars or black holes,\footnote{
Another tempting issue is whether the M-matter itself 
could help in producing big central black holes, 
with masses $\sim 10^7~ M_\odot$, which are thought 
to be main engines of the active galactic nuclei.
} and reproducing the mirror gas and dust.  
Although the cross section $\sigma_H$ is large, 
it does not necessarily implies that the galaxy core 
will collapse within a dynamical time, since the inner 
halo should be opaque for M-particles. 
They undergo many scatterings and escape from the 
system via diffusion, so the energy drain can be 
small enough and the instability time can substantially 
exceed the age of the universe \cite{hannestad}. 

In the galactic halo 
(provided that it is the elliptical mirror galaxy) 
the mirror stars should be observed as 
Machos in gravitational microlensing  \cite{BDM,Macho}.  
Leaving aside the difficult question of the initial 
stellar mass function, one can remark that once  
the mirror stars could be very old 
and evolve faster than the ordinary ones, 
it is suggestive to think that most  
of massive ones, with mass above the 
Chandrasekhar limit $M_{\rm Ch} \simeq 1.5 ~ M_\odot$ 
have already ended up as supernovae, so that only the 
lighter ones 
remain as the microlensing objects.\footnote{The 
M-supernovae explosion in our galaxy cannot be directly 
seen by ordinary observer, however it  
could be observed in terms of gravitational waves. 
In addition, if the M- and O-neutrinos are mixed \cite{FV,BM}, 
it can lead the observable neutrino signal, which could 
be also accompanied by the weak gamma ray burst \cite{GRB}. 
}
The recent data indicate the average mass of 
Machos around $M\simeq 0.5 ~M_\odot$, which is difficult 
to explain in terms of the brown dwarves with masses 
below the hydrogen ignition limit $M < 0.1 M_{\odot}$ 
or other baryonic objects \cite{Freese}. 
Perhaps, this is the first observational evidence 
of the mirror matter? 

It is also plausible that in the galactic halo 
some fraction of mirror stars exists in the form 
of compact substructures like globular or open clusters. 
In this case, for a significant statistics, one could 
observe interesting time and angular correlations 
between the microlensing events.

\section{Conclusions}

We have discussed cosmological implications of the 
parallel mirror world with the same microphysics 
as the ordinary one, but having smaller temperature, 
$T'< T$, with the limit $T'/T< 0.64$  set by 
the BBN constraints.   
Therefore, the M-sector contains less relativistic 
matter (photons and neutrinos) than the O-sector,  
$\Omega'_r \ll \Omega_r$. 
On the other hand, in the context of the GUT or 
electroweak baryogenesis scenarios, the condition 
$T'<T$ yields that the mirror sector should produce a 
larger baryon asymmetry 
than the observable one, $\eta'_B>\eta_B$.\footnote{
In some sense, this is true also in the case of the 
Affleck-Dine mechanism \cite{Dolgov}. Namely, 
one can show that for the same initial values of the 
baryonic charge breaking fields in two sectors, 
the larger baryon asymmetry will occur in the sector 
which has the lower temperature. This question 
will be discussed in more details elsewhere.
}
So, in the relativistic expansion epoch the cosmological 
energy density is dominated by the ordinary component,  
while the mirror one gives a negligible contribution.  
However, for the non-relativistic epoch  
the complementary situation can occur when  
the mirror baryon matter density is bigger 
than the ordinary one, $\Omega'_B > \Omega_B$. 
Hence, the MBDM can contribute 
the dark matter of the universe along with the CDM  
or even constitute a dominant dark matter component.     

We have also shown that the BBN epoch in the mirror 
world proceeds differently from the ordinary one, 
and it predicts the mirror helium abundance in the 
range $Y'_4 =0.4-0.8$, considerably larger than 
the observable $Y_4\simeq 0.24$. 
 
Unfortunately, we cannot exchange the information 
with the mirror physicists and combine our observations. 
(Afterall, since two worlds have the same microphysics, 
the life should be possible also in the mirror sector.)    
However, there can be many possibilities to disentangle 
the cosmological scenario of two parallel worlds 
with the the future high precision data concerning 
the large scale structure, CMB anisotropy, 
structure of the galaxy halos, gravitational 
microlensing, oscillation of the neutrinos or other 
neutral particles into their mirror partners, etc.

\section*{Acknowledgements}

We thank to Venya Berezinsky, Silvio Bonometto, 
Sasha Dolgov, Andrei Doroshkevich, Victor Dubrovich, 
Gianni Fiorentini, Lev Kofman,  Alex Melchiorri 
and Subir Sarkar for useful discussions. 
The work is partially supported by the MURST 
research grant "Astroparticle Physics".

\end{document}